\documentstyle{article}

\newcommand{\lag}{{\mathcal{L}}}

\newcommand{\ltsim}{\lower3pt\hbox{$\, \buildrel < \over \sim \, $}}
\newcommand{\gtsim}{\lower3pt\hbox{$\, \buildrel > \over \sim \, $}}
\newcommand{\glt}{\lower3pt\hbox{$\, \buildrel < \over > \, $}}

\renewcommand{\d}{{\mathrm{d}}}
\newcommand{\e}{{\mathrm{e}}}

\begin{document}

\rightline{UG--FT--124/00}
\rightline{hep-ph/0011143}
\rightline{November 2000}

\vspace{0.5cm}
\begin{center}

\large {\bf Large top mixing from extra dimensions}
\vspace*{5mm}
\normalsize

{\bf F. del Aguila} and {\bf J. Santiago}

\smallskip 
\medskip 
{\it Departamento de F\'{\i}sica Te\'{o}rica y del Cosmos} \\
{\it Universidad de Granada, E-18071 Granada, Spain}
\smallskip  

\vskip0.2in \end{center}

\begin{abstract}
Fermion mixing is conveniently described  using the effective
Lagrangian formalism. We apply this approach to study top mixing in
models with an infinite tower of Kaluza-Klein fermion excitations. In
the Randall-Sundrum background with a boundary Higgs and
 phenomenologically viable values
of the model parameters, the only effect eventually
observable is the loss of universality of the top couplings. Their
deviation from the SM predictions can be up to $4\%$ if the
five-dimensional Yukawa
couplings times the ${\mathrm AdS_5}$ curvature scale are $\leq 10$.
\end{abstract}
\footnotetext[1]{Presented by F.A. at NATO Advanced Study Institute 2000
``Recent Developments in Particle Physics and Cosmology'', Cascais 26
June--7 July 2000.}\addtocounter{footnote}{1}
\section{Introduction}

Fermion mixing has been historically a window to new physics and a
guide for model building. A primer example was the prediction of the
existance of the charm quark to explain the suppression of
strangeness-changing neutral currents (GIM
mechanism)~\cite{gim}. Similarly the naturalness of such a suppression
for all Flavour-Changing Neutral Currents (FCNC) implies that the
isospin properties of the three standard families are the
same~\cite{glashowweinberg:77}. However, this may be not what is
required to describe
the top quark. Its mixing is poorly known experimentally,
leaving enough room for relatively large deviations from the Standard
Model (SM) predictions (see Ref.~\cite{beneke} for a review). As a
matter of fact, there are simple SM extensions where FCNC not
involving the top quark are naturally suppressed in a large region of
parameter space, and which allow for a large top mixing at the same 
time~\cite{aguila:ja:miquel}. In the following,  we first 
discuss the limits on top mixing~\cite{beneke} and the description of 
fermion mixing using the effective Lagrangian
approach~\cite{apvs1}. Then we apply the general results for heavy
exotic fermions~\cite{apvs2}
to top mixing in the five-dimensional Randall-Sundrum (RS)
model~\cite{rs,rsphenomenology,gherghetta,davoudiasl:all,as1,huber:shafi:1} 
and show that the size of possible SM deviations can be up to $4\%$.

\section{Top quark couplings}

The lowest dimension gauge interactions involving the top quark can be
written in standard notation~\cite{beneke,pdb,zeuthen}

\begin{eqnarray}
\label{lag4:vtq}
{\mathcal{L}}^{Vtq}_4&=& -g_s \bar{t} \gamma^\mu T^a t G_{\mu
a}-\frac{2}{3}e\bar{t}\gamma^\mu t A_\mu \\
& &-\frac{g}{\sqrt{2}}\sum_{q=d,s,b}\bar{t}\gamma^\mu
(v^W_{tq}-a^W_{tq}\gamma_5)q W^+_\mu + \mathrm{h.c.} \nonumber \\
& & -\frac{g}{2\cos \theta_W} \bar{t}  \gamma^\mu
(v^Z_{tt}-a^Z_{tt}\gamma_5)t Z_\mu \nonumber \\
& &-\frac{g}{2\cos\theta_W}\sum_{q=u,c}\bar{t}\gamma^\mu
(v^Z_{tq}-a^Z_{tq}\gamma_5)q Z_\mu + \mathrm{h.c.}. \nonumber
\end{eqnarray}
The first two terms are fixed by the unbroken gauge symmetry
$SU(3)_C\times U(1)_Q$, while 
in the SM the charged currents are parametrized by
$v^W_{tq},a^W_{tq}=\frac{V_{tq}}{2}$, 
where $V_{tq}$ is the Cabibbo-Kobayashi-Maskawa (CKM) matrix
\cite{ckm}, and the neutral currents by
$v^Z_{tt}=\frac{1}{2}-\frac{4}{3}\sin^2\theta_W
$, $a^Z_{tt}=\frac{1}{2}$ and
$v^Z_ {tq},a^Z_{tq}=0$. The SM Yukawa couplings are also diagonal. For
discussing possible SM extensions using the effective Lagrangian
formalism above the electroweak scale 
it is more convenient, however, to work with left (LH) and right 
(RH)-handed fields. Thus 
\begin{eqnarray}
\lag^{Vqq^\prime}_4&=&
-\frac{g}{\sqrt{2}}(
\bar{u}^i_L W^{L}_{ij}
\gamma^\mu d^j_L+
\bar{u}^i_R W^{R}_{ij}
\gamma^\mu d^j_R)W^+_\mu +\textit{h.c.}
\nonumber \\
&&-\frac{g}{2\cos \theta_W} \left( 
\bar{u}^i_L X^{uL}_{ij}
\gamma^\mu u^j_L+
\bar{u}^i_R X^{uR}_{ij}
\gamma^\mu u^j_R \right.\nonumber \\
&&\left.-\bar{d}^i_L X^{dL}_{ij}
\gamma^\mu d^j_L-
\bar{d}^i_R X^{dR}_{ij}
\gamma^\mu d^j_R - 2\sin^2\theta_W J^\mu_{\mathrm{EM}}\right) Z_\mu,
 \label{lag:ZWH}
\end{eqnarray}
where the summation on family indices $i,j=1,2,3$ is understood and 
\begin{equation}
\begin{array}{ll}
v^W=\frac{W^L+W^R}{2}, & a^W=\frac{W^L-W^R}{2}, \\
v^Z=\frac{X^L+X^R}{2}-2\sin^2\theta_W Q, & a^Z=\frac{X^L-X^R}{2}, 
\end{array}
\end{equation}
with the electric charge $Q$ term only substracted for the diagonal
couplings.  

Present limits on FCNC involving the
light quarks are somewhat stringent, but this is not yet the case for 
the top quark~\cite{beneke,pdb,lep:singletop}. For instance,
from the size of $K^+ \to \pi^+
\nu \bar{\nu}$ and neglecting large cancellations with four-fermion
contributions $|X^{L,R}_{ds}| < 4 \times 10^{-5}$, whereas available collider
data only imply $|X^{L,R}_{tq}|< 0.66$.
 Top couplings are constrained by electroweak data within the SM with a higher
precision than will be by 
future direct measurements at large colliders. The
question is if one can expect to observe departures
from the SM top predictions in future experiments. 
 The answer is positive as we will discuss 
in last section.

\section{Top quark effective Lagrangian}

\begin{table*}[!t]
\caption{Dimension 6 operators contributing to renormalizable $Vtq$
couplings after SSB.}
\label{operators:tab}\vspace{0.5cm}
\begin{tabular}{lll}
${\mathcal{O}}_{\phi q}^{(1)}= 
(\phi^\dagger i
D_\mu\phi)(\bar{q}\gamma^\mu q)$
&
${\mathcal{O}}_{u \phi}= 
(\phi^\dagger \phi)(\bar{q} u \tilde{\phi})$
&${\mathcal{O}}_{\phi W}=\frac{1}{2}(\phi^\dagger
\phi)W^I_{\mu\nu}W^{I\mu\nu}  $
\\
${\mathcal{O}}_{\phi q}^{(3)}= 
(\phi^\dagger \tau^I i
D_\mu\phi)(\bar{q}\gamma^\mu \tau^I q)$
& 
${\mathcal{O}}_{d \phi}= 
(\phi^\dagger \phi)(\bar{q} d \phi)$
&
${\mathcal{O}}_{\phi B}=\frac{1}{2}(\phi^\dagger
\phi)B_{\mu\nu}B^{\mu\nu} $
\\ 
 ${\mathcal{O}}_{\phi u}= (\phi^\dagger i
D_\mu\phi)(\bar{u}\gamma^\mu u)$
&
&
${\mathcal{O}}_{W B}=(\phi^\dagger \tau^I
\phi)W^I_{\mu\nu}B^{\mu\nu} $\\ 
 ${\mathcal{O}}_{\phi d}= (\phi^\dagger i
D_\mu\phi)(\bar{d}\gamma^\mu d)$
& &
${\mathcal{O}}_\phi^{(1)}=(\phi^\dagger \phi)  (D_\mu\phi^\dagger
D^\mu\phi)  $\\ 
${\mathcal{O}}_{\phi \phi}= (\phi^T \epsilon i
D_\mu\phi)(\bar{u}\gamma^\mu d)$
& &
${\mathcal{O}}_\phi^{(3)}=(\phi^\dagger D^\mu\phi)  (D_\mu\phi^\dagger
\phi) $ 
\end{tabular}
\vspace{0.5cm}
\end{table*}

The unbroken $U(1)_Q$ protects the term proportional to
$J^\mu_{\mathrm{EM}}$ in Eq. (\ref{lag:ZWH}). The other Z and W
couplings get contributions from higher orders in the effective
Lagrangian expansion after the Spontaneous Symmetry Breaking (SSB) of 
$SU(2)_L\times U(1)_Y$. The lowest order corrections result from the
dimension 6 Lagrangian~\cite{buchmuller}
\begin{equation}
\lag_6^{\mathit eff}=\frac{\alpha_x}{\Lambda^2} {\mathcal O}_x +
{\mathrm h.c.},
\end{equation} 
where $\Lambda$ is the effective high scale and the relevant operators
${\mathcal O}_x$
are collected in Table~\ref{operators:tab}. As a function of the
coefficients $\alpha_x$ the $X$ and $W$ coupling matrices in
Eq. (\ref{lag:ZWH}) read to order $\frac{v^2}{\Lambda^2}$, with $v$
the electroweak vacuum expectation value,~\cite{apvs1}
\begin{eqnarray}
X^{uL}_{ij}&=& \delta_{ij}-\frac{1}{2}\frac{v^2}{\Lambda^2}
V_{ik}(\alpha^{(1)}_{\phi q}+\alpha^{(1)\dagger}_{\phi q}
-\alpha^{(3)}_{\phi q}-\alpha^{(3)\dagger}_{\phi q}
)_{kl}V^\dagger_{lj}, \nonumber \\
X^{uR}_{ij}&=& -\frac{1}{2}\frac{v^2}{\Lambda^2}
(\alpha_{\phi u}+\alpha_{\phi u}^\dagger
)_{ij}, \nonumber \\
X^{dL}_{ij}&=& \delta_{ij}+\frac{1}{2}\frac{v^2}{\Lambda^2}
(\alpha^{(1)}_{\phi q}+\alpha^{(1)\dagger}_{\phi q}+
\alpha^{(3)}_{\phi q}+  
\alpha^{(3)\dagger}_{\phi
q})_{ij}, \nonumber \\
X^{dR}_{ij}&=& \frac{1}{2}\frac{v^2}{\Lambda^2}
(\alpha_{\phi d}+\alpha_{\phi d}^\dagger
)_{ij},  \label{couplings}\\
W^{L}_{ij}&=& \tilde{V}_{ik}\left(\delta_{kj}+\frac{v^2}{\Lambda^2}
(\alpha^{(3)}_{\phi q})_{kj}\right), \nonumber \\
W^{R}_{ij}&=& -\frac{1}{2}\frac{v^2}{\Lambda^2}
(\alpha_{\phi \phi})_{ij},\nonumber
\end{eqnarray}
where
\begin{equation}
\tilde{V} = V+ \frac{v^2}{\Lambda^2} (V A_L^d - A_L^u V) \, 
\end{equation}
is a unitary matrix redefining the initial CKM matrix $V$ to take into
account the diagonalization of the quark mass matrices to order
$v^2/\Lambda^2$. $A^{u,d}_L$ only involve the coefficients of the
dimension 6 operators ${\mathcal O}_{u\phi,d\phi}$, respectively. The
non-linear realization of this Lagrangian
is studied in Ref.~\cite{espriu:manzano}.

The couplings in Eq. (\ref{couplings})  incorporate features that are
forbidden in the SM, namely, FCNC,
RH neutral currents not proportional to $J^\mu_{\mathrm{EM}}$,
RH charged currents, and LH charged currents which
are not described by a unitary matrix. 
These couplings can be
directly determined for instance from processes involving the $V \bar{q}
q^{\prime}$ vertices in which the final
$V$ is 
observed. For the top 
quark this will be possible at large
colliders~\cite{beneke,delAguila:2000ec}.  
 Trilinear couplings also contribute to four-fermion
processes (such as mixing of neutral mesons), but
four-fermion operators may contribute in this case as
well. One can still use these processes to put limits on these
couplings assuming that there are no strong cancellations between
cubic and quartic couplings.

The operators in the last column of Table~\ref{operators:tab} redefine
the gauge bosons and thus  the trilinear couplings in
Eq. (\ref{couplings}). However, they are flavour blind and hence
also constrained by precise electroweak data involving only the light
quark flavours, giving then unobservable
corrections to top couplings.

\section{Bulk fermions in the RS model}

Models in extra dimensions can give towers of Kaluza--Klein (KK)
excitations in four dimensions not far from the TeV
scale~\cite{antoniadis}. The integration of KK gauge bosons and/or
fermions generate $v^2/\Lambda^2$ corrections to
the $X$ and $W$ couplings in Eq. (\ref{couplings}). In both cases
they can be originated at tree level and can be {\itshape a priori}
 large~\cite{arzt}. 
 However, the KK gauge boson contributions are proportional to their
couplings to the SM fermions times their couplings to
the SM gauge bosons, and the latter are typically $\sim
0.01$~\cite{pdb,Langacker}. In this case the corresponding top corrections
would be too small to be observable. In the following we will
concentrate on the KK fermion contributions, studying in detail the
case of bulk fermions in the RS model~\cite{as1}.

In five dimensions there are no chiral fermions. Thus five-dimensional 
fermions $\Psi$ are vector-like and can have a Dirac mass of the form 
\begin{equation}
\lag_{\mathrm D}
=- i m_\Psi (\bar{\Psi}_{\mathrm L} 
\Psi_{\mathrm R} +
\bar{\Psi}_{\mathrm R}
 \Psi_{\mathrm L}),
\label{lag:dirac}
\end{equation}
where $\Psi$ is the sum of the two four-dimensional ``chiralities'' 
$\Psi_{\mathrm{L,R}}=\pm \gamma_5 \Psi_{\mathrm{L,R}}$, 
transforming in the same way under the gauge group.
 The five-dimensional action containing the 
Yukawa interactions for the three standard families 
can  be written in general with the Higgs on the TeV ($\pi R$) brane 
\begin{eqnarray}
S_{\mathrm Yuk}&=&-i
\int \d^4x\;\int \d y\;\sqrt{-g}\left[ \lambda^{u(5)}_{ij} \bar{q}_i(x,y)
\tilde{u}_j(x,y) \tilde{\phi}(x)   \right. \nonumber \\
&&+ \left.  \lambda^{d(5)}_{ij} \bar{q}_i(x,y)
\tilde{d}_j(x,y) \phi(x) + {\mathrm h.c.}
\right] \delta(y-\pi R).\label{S:yuk}
\end{eqnarray} 
Expanding the five-dimensional fields in KK towers and 
integrating over the fifth dimension,  
 one obtains the
four-dimensional mass Lagrangian which after SSB reads
\begin{eqnarray}
i\lag_{\mathrm mass}&=&\sum_{n,m=0}^\infty\left[ 
\lambda^{u(nm)}_{ij}
\bar{u}^{(n)i}_L\tilde{u}^{(m)j}_R
+\lambda^{d(nm)}_{ij}
\bar{d}^{(n)i}_L\tilde{d}^{(m)j}_R
\right] +{\mathrm h.c.} \nonumber \\
&& +\sum_{n=1}^\infty \left[
M^{q(n)}_i
( \bar{u}^{(n)i}_L u^{(n)i}_R+\bar{u}^{(n)i}_R u^{(n)i}_L+
\bar{d}^{(n)i}_L d^{(n)i}_R+\bar{d}^{(n)i}_R d^{(n)i}_L) \right.
 \nonumber \\
&&\quad \quad
+M^{u(n)}_i (\bar{\tilde{u}}^{(n)i}_L \tilde{u}^{(n)i}_R +\bar{\tilde{u}}^{(n)i}_R
\tilde{u}^{(n)i}_L) \nonumber \\
&& \quad \quad
\left.+M^{d(n)}_i (\bar{\tilde{d}}^{(n)i}_L \tilde{d}^{(n)i}_R
+\bar{\tilde{d}}^{(n)i}_R 
\tilde{d}^{(n)i}_L)\right],\label{lag:mass}
\end{eqnarray}
where we have added to Eq. (\ref{S:yuk}) the corresponding
Dirac masses in Eq.
(\ref{lag:dirac}). The latter can always be taken diagonal.
 The four-dimensional masses can be written
\begin{eqnarray}
\lambda^{u,d(nm)}_{ij}&=&\lambda^{u,d}_{ij} a_q^{(n)i} a_{u,d}^{(m)j}, 
\end{eqnarray}
with
\begin{eqnarray}
\lambda^{u,d}_{ij}&=&\lambda^{u,d(5)}_{ij} k \frac{v}{\sqrt{2}}
\sim {\mathrm \;SM\; masses}, \nonumber \\
a_q^{(n)i} &=& \e^{\pi k R/2} \frac{f_{qL}^{(n)i}(\pi R)}{\sqrt{2\pi k
R}}, \\
a_{u,d}^{(m)j}&=& \e^{\pi k R/2} \frac{f_{u,dR}^{(m)j}(\pi
R)}{\sqrt{2\pi k R}}, \nonumber
\end{eqnarray}
and
\begin{eqnarray} 
v&=&\e^{-\pi k R} v^{(5)}\sim 250 \; {\mathrm GeV}.
\end{eqnarray}
The factor
 $\e^{\pi k R}$ in  $\lambda^{u,d(nm)}_{ij}$
is due to the rescaling of the boundary Higgs 
canonically normalized and the expansion coefficients $f^{(n)}$ give
 the {\it n}-th fermion wave functions at the $\pi {\mathrm R}$ brane
 ($k\sim 2.44\times 10^{18}$ GeV and $kR\sim 11$~\cite{huber:shafi:1}). 
 It must be noticed that 
 odd fields are zero at the TeV boundary and then the odd
chiralities ($q_{\mathrm R},\tilde{u}_{\mathrm L},\tilde{d}_{\mathrm L}
$) have zero Yukawa couplings for a boundary Higgs.
In matrix notation Eq. (\ref{lag:mass}) reads
\begin{equation}
{\mathcal{M}}^u=
\begin{array}{l}
 \\ \bar{u}^{(0)}_L \\\bar{\tilde{u}}^{(1)}_L   \\ \vdots \\
\bar{u}^{(1)}_L \\ \vdots
\end{array}
\begin{array}{c} \quad \quad 
  \tilde{u}^{(0)}_R  \quad \quad
\tilde{u}^{(1)}_R  \quad \quad \quad \ldots  \quad
u^{(1)}_R  \quad  \ldots \\
\left(
\begin{array}{ccccc}
 \lambda^u_{ij} a^{(0)i}_q a^{(0)j}_u 
 & \lambda^u_{ij} a^{(0)i}_q a^{(1)j}_u 
&  \ldots & 0 & \dots \\
 0 & M^{u(1)}_i \delta_{ij}
 & \ldots & 0  & \dots \\
 \vdots & \vdots
 & \ddots & \vdots &  \\
 \lambda^u_{ij} a^{(1)i}_q a^{(0)j}_u  & \lambda^u_{ij} a^{(1)i}_q
 a^{(1)j}_u   
 & \ldots & M^{q(1)}_i \delta_{ij}  & \dots \\
 \vdots & \vdots &  & \vdots  & \ddots
\end{array}\right) , \end{array}
\end{equation}
and similarly for ${\mathcal{M}}^d$. It is finally convenient in order
to compare with experiment, to rotate the zero modes
\begin{equation}
\tilde{u}^{(0)i}_R=(U^u_R)_{ij} \tilde{u}^{\prime(0)j}_R, \quad 
\tilde{d}^{(0)i}_R=(U^d_R)_{ij} \tilde{d}^{\prime(0)j}_R,\quad 
q^{(0)i}_L=(U^q_L)_{ij} q^{\prime(0)j}_L,
\end{equation}
where $\tilde{u}_R^{\prime (0)},\;\tilde{d}_R^{\prime
(0)},\;q_L^{\prime (0)}$ are the quark mass 
eigenstates up to mixing with the
heavy KK excitations. In this 
basis the $3\times 3$ light mass submatrices
are  
\begin{equation}
(U^{q\dagger}_{L})_{ik} \lambda^u_{kl} a^{(0)k}_q a^{(0)l}_u
(U^u_R)_{lj} 
=
V^\dagger_{ij} m^u_j,\quad
(U^{q\dagger}_{L})_{ik} \lambda^d_{kl} a^{(0)k}_q a^{(0)l}_d 
(U^d_R)_{lj}  =
m^d_i \delta_{ij}, \label{diagonalization}
\end{equation}
with $m^{u,d}_i$ the quark masses and $V$ 
the CKM matrix in the absence of mixing. The effective Lagrangian
resulting from integrating out the heavy vector-like KK fermions is
explicitly given in Refs.~\cite{apvs2,as1}. For the RS model the effective 
couplings in Eq. (\ref{couplings}) read 
\begin{eqnarray}
X^{uL}_{ij}&=&\delta_{ij}-
m^u_i (U^{u\dagger}_R)_{ik}
\left[\sum_{n=1}^\infty 
\left(\frac{a^{(n)k}_u}{a^{(0)k}_u}\right)^2 
\frac{1}{M^{u(n)\;2}_k}
\right]
 (U^{u}_R)_{kj} m^u_j , 
\label{xul1} \nonumber \\
X^{uR}_{ij}&=&
m^u_i V_{il} 
(U^{q\dagger}_L)_{lk}
\left[\sum_{n=1}^\infty 
\left(\frac{a^{(n)k}_q}{a^{(0)k}_q}\right)^2 
\frac{1}{M^{q(n)\;2}_k}
\right]
 (U^{q}_L)_{kr} V^\dagger_{rj} m^u_j , 
\label{xur1} \nonumber \\
X^{dL}_{ij}&=&\delta_{ij}-
m^d_i (U^{d\dagger}_R)_{ik}
\left[\sum_{n=1}^\infty 
\left(\frac{a^{(n)k}_d}{a^{(0)k}_d}\right)^2 
\frac{1}{M^{d(n)\;2}_k}
\right]
 (U^{d}_R)_{kj} m^d_j ,
\label{xdl1}  \\
X^{dR}_{ij}&=&
m^d_i  (U^{q\dagger}_L)_{ik}
\left[\sum_{n=1}^\infty 
\left(\frac{a^{(n)k}_q}{a^{(0)k}_q}\right)^2 
\frac{1}{M^{q(n)\;2}_k}
\right]
 (U^{q}_L)_{kj}  m^d_j , 
\label{xdr1} \nonumber \\
W^{L}_{ij}&=&\tilde{V}_{ij}
-\frac{1}{2}
m^u_i (U^{u\dagger}_R)_{ik}
\left[\sum_{n=1}^\infty 
\left(\frac{a^{(n)k}_u}{a^{(0)k}_u}\right)^2 
\frac{1}{M^{u(n)\;2}_k}
\right]
 (U^{u}_R)_{kl} m^u_l
\tilde{V}_{lj} \nonumber \\
&&-\frac{1}{2}
\tilde{V}_{il}
m^d_l (U^{d\dagger}_R)_{lk}
\left[\sum_{n=1}^\infty 
\left(\frac{a^{(n)k}_d}{a^{(0)k}_d}\right)^2 
\frac{1}{M^{d(n)\;2}_k}
\right]
 (U^{d}_R)_{kj} m^d_j , 
\label{wl1} \nonumber \\
W^{R}_{ij}&=&
m^u_i V_{il} (U^{q\dagger}_L)_{lk}
\left[\sum_{n=1}^\infty 
\left(\frac{a^{(n)k}_q}{a^{(0)k}_q}\right)^2 
\frac{1}{M^{q(n)\;2}_k}
\right]
 (U^{q}_L)_{kj}  m^d_j , 
\label{wr1} \nonumber
\end{eqnarray}
where at this order $V$ can be replaced by $\tilde V$ in 
$X^{uR}$ and $W^R$.
 The extra contributions are
products of 
$3\times 3$ matrices, where the one in 
square brackets is diagonal and depends on the fermion location and
then on the details of 
the model, and the others are unitary 
combinations of SM masses.  
The matrix in the middle can be further simplified 
noting that
\begin{equation}
(a^{(n)})^2=(a^{(1)})^2=1, 
\end{equation}
which, up to a constant, leaves the diagonal elements 
as an infinite sum of the inverse of the KK heavy masses $M^{(n)}$ 
squared.  The lightest mass $M^{(1)}$ plays the r\^ole of the
effective scale in Eq. (\ref{couplings}), and the SM masses include
 the electroweak
vacuum expectation value. The size of
the corrections in the RS background
depends on one mass parameter per flavour and
$SU(2)_{\mathrm L}\times U(1)_{\mathrm Y}$ multiplet. One can try to
explain the observed pattern of fermion masses in this extended
model~\cite{huber:shafi:2} or as we do in the following, simply ask 
how large can the SM corrections to the effective
couplings be.

\begin{table}[b]
\begin{center}
\caption{Experimental limits expected at LHC for the top quark flavour
changing branching ratios ${\mathrm Br(t\to uZ,cZ)}$ and values
predicted in
the SM, the two Higgs model (2H), supersymmetric models (SUSY) without
R and with $\not \!  \! R$   parity breaking and the SM extensions
with  exotic quarks. The branching ratio is defined as
the decay rate divided by 1.56  GeV.
}
\label{br:models}\vspace{0.5cm}
\begin{tabular}{ccccc}
LHC&SM&2H& $\begin{array}{c} {\mathrm SUSY} \\
{\mathrm \not{\!\mbox{R}},R}\end{array}$  & $\begin{array}{c}
{\mathrm Vector-like} \\ {\mathrm quarks} \end{array}$ \\
\hline
$\sim 10^{-4}$ & $\sim 10^{-13}$  & $\sim10^{-6}$ & $\sim10^{-4},10^{-8}$ 
 & $\sim 10^{-2}$  
\end{tabular}
\vspace{0.5cm}
\end{center}
\end{table}

\section{Experimental implications}

In popular SM extensions the expected flavour changing branching
ratios for the top quark are too
small ($<10^{-4}$) to be observable at future colliders, except in the
case of extra vector-like quarks near the electroweak scale (see
Table~\ref{br:models}~\cite{beneke}). For SM extensions with exotic
quarks~\cite{aguila:bowick} electroweak data imply for instance
$|X^L_{tc}|<0.08$ and $|X^R_{tc}|<0.16$~\cite{aguila:ja:miquel}, and
then branching ratios $\ltsim 10^{-2}$. The RS model which is a
particular case with an infinite tower of vector-like quark singlets
and doublets must satisfy these limits. In fact, the fit of
Eq. (\ref{xdl1}) to present data saturates these bounds. However 
for $M^{(1)}\gtsim10$ TeV~\cite{huber:shafi:1} the corresponding Yukawa
couplings $\lambda^{(5)} k$ 
are $\sim 100$ and thus too large to recover the successful
SM description below the TeV scale. Allowing only for
$\lambda^{(5)}k\leq 10$ the top coupling
corrections are reduced, becoming only eventually
observable for $X_{tt}$ and
$W^L_{tb}$. In this region of parameters $|\Delta X_{ij}|\sim
\frac{m_i m_j}{m^2_t}\Delta X^{uL}_{tt}$ and then the main
phenomenological constraints result from the loss of universality of
the top couplings~\cite{as1}. Maximizing the top mixing in this region
of parameter space and requiring at least the same
$\chi^2$ as for the SM we have obtained departures from the SM top
couplings of up to $\sim 4\%$. Indeed for the Yukawa couplings
\begin{equation}
\lambda^{u(5)}_{ij} k=\left(
\begin{array}{ccc}  
5.6\times 10^{-4} & 10 & 9.2\times 10^{-3} \\
-6.2\times 10^{-4} & -10 & -9.2\times 10^{-3} \\
1.2\times 10^{-3} & -10 & 1.8\times 10^{-2} \\
\end{array}
\right),
\end{equation}
\begin{equation}
\lambda^{d(5)}_{ij} k=\left(
\begin{array}{ccc}  
-5.9\times 10^{-5} & -1.1\times10^{-3} & 4.2\times 10^{-2} \\
-7.7\times 10^{-5} & 5.2\times10^{-4} & -4.2\times 10^{-2} \\
1.8\times 10^{-5} & -1.5\times10^{-3} & -4.6\times 10^{-2} \\
\end{array}
\right)
\end{equation}
and the coefficients 
\begin{eqnarray}
a_{u}^{(0)u,c,t}&=&0.5603,0.1018,0.5603,\nonumber \\
a_d^{(0)i}&=&
a_q^{(0)i}=0.5603, \label{as}
\end{eqnarray}
where $i=1,2,3$ is the family index, 
 we find 
\begin{equation}
X^{L}_{tt}=0.9608,\quad X^{R}_{tt}=0.0013,
\end{equation}
and
\begin{equation}
|W^L_{ij}|=\left(
\begin{array}{ccc}
0.9752 & 0.2227 & 0.0036 \\
0.2227 & 0.9752 & 0.0402 \\
0.0096 & 0.0394 & 0.9804 
\end{array}
\right).
\end{equation}
The corrections to the remaining $X$ and $W$ 
couplings are negligible. We have
 taken $k=2.44\times 10^{18}$ and $k R=10.815$~\cite{huber:shafi:1}.
 The rotation matrices $U$ entering in Eq. (\ref{xdl1}) are 
fixed by the standard fermion masses 
and the CKM matrix through Eq. (\ref{diagonalization}).
 Note that the value $a^{(0)}=0.5603$ in Eq. (\ref{as})  ensures 
that the bound on the lightest KK gauge boson mass 
$M_1^{\mathrm gauge}\gtsim 10$ TeV is satisfied. This is needed to
suppress the contribution of the tower of KK gauge bosons to electroweak
 observables below present experimental limits~\cite{davoudiasl:all}.

In summary, we have shown that in the RS model there can be
deviations from the SM
$Z\bar{t}t$ and $W\bar{t}b$ couplings of at most $\sim 4\%$ and $2\%$, 
respectively, due to the top mixing with the KK fermion
excitations. This has to be compared with the
measurement of $W^L_{tb}$ at LHC, for which an accuracy of $5\%$
($10\%$ for $|W^L_{tb}|^2$ in the cross section) is forseen as an
ambitious but attainable goal~\cite{beneke}. There are better
prospects for the measurement of $X^L_{tt}$ at TESLA. For instance,
with a center of mass energy of 500 GeV, an integrated luminosity 
of 300 fb$^{-1}$ and unpolarized beams
one expects to collect in the detector 34800  top pairs   
with one $W$ decaying into $e\nu$ or $\mu\nu$ and the
 other $W$ decaying
hadronically, reaching 
a precision of $2\%$ in the determination
of the $Z\bar{t}t$ coupling~\cite{ja:private}. This precision will
improve when all channels are included.

\section*{Acknowledgements}

This work has been supported by CICYT and Junta de Andaluc\'{\i}a. We
thank J.A. Aguilar-Saavedra and M. P\'erez-Victoria 
for useful comments. F.A. also
thanks the organizers of the School for their hospitality and J.S.
thanks MECD for financial support.

\end{document}